\documentclass[prb,twocolumn,aps,showpacs]{revtex4}
\usepackage{graphics,graphicx,epsfig}
\begin{document}
\title{Kinetic roughening and porosity scaling in film growth with
subsurface lateral aggregation}
\author{F. D. A. Aar\~ao Reis\footnote{Email address: reis@if.uff.br}}
\affiliation{
Instituto de F\'\i sica, Universidade Federal Fluminense,\\
Avenida Litor\^anea s/n, 24210-340 Niter\'oi RJ, Brazil}
\date{\today}

\begin{abstract}

We study surface and bulk properties of porous films produced by a model
in which particles incide perpendicularly to a substrate, interact with
deposited neighbors in its trajectory, and aggregate laterally with
probability of order $a$ at each position. The model generalizes
ballistic-like models by allowing attachment to particles below the outer
surface. For small values of $a$, a crossover from uncorrelated deposition
(UD) to correlated growth is observed.
Simulations are performed in $1+1$ and $2+1$ dimensions.
Extrapolation of effective exponents and comparison of roughness
distributions confirm Kardar-Parisi-Zhang roughening of the outer surface
for $a>0$.
A scaling approach for small $a$ predicts crossover times as $a^{-2/3}$
and local height fluctuations as $a^{-1/3}$ at the crossover,
independently of substrate dimension.
These relations are different from all previously studied models
with crossovers from UD to correlated growth due to subsurface
aggregation, which reduces scaling exponents.
The same approach predicts the porosity and average pore height
scaling as $a^{1/3}$ and $a^{-1/3}$, respectively,
in good agreement with simulation results in $1+1$ and $2+1$ dimensions.
These results may be useful to modeling samples with
desired porosity and long pores.

\end{abstract}

\pacs{81.15.Aa, 05.40.-a, 68.35.Ct, 68.55.-a}
\maketitle

\section{Introduction}
\label{intro}

A widely studied model for growth of porous films is ballistic deposition (BD) \cite{vold},
in which the particles incide perperdicularly
to the substrate and aggregate at the first contact with the deposit \cite{barabasi,krug}.
BD was originally proposed to describe sedimentary rock formation \cite{vold}
and was extended to model thin film growth and related systems by considering other
aggregation mechanisms, non-colimated particle flux, or polydispersivity of particle size
\cite{albano1,katzav2004,hivert,juvenil,perez,yanguas,ye,lehnen,bbd1}.
Most works on the ballistic-like models address the scaling features of the outer
surface of the deposits,
particularly for the connections with Kardar-Parisi-Zhang (KPZ) \cite{kpz} roughening.
Some works also connect the surface growth dynamics and the bulk properties of
the porous deposits \cite{strucbelow,yu,bbdflavio,khanin,juvenil,tang2013}.
This is an essential step for proposing models of porous materials,
which have a large variety of technological applications, frequently in the
form of thin films \cite{innocenzi,shekhah}.

Some ballistic-like models are in a class of competitive growth models in which
uncorrelated deposition (UD) is obtained for a certain value of a parameter
and, near that value, a crossovers in kinetic roughening
is observed \cite{albano1,juvenil,bbd1}.
In the simplest model, particle aggregation follows the BD (UD) rule with probability
$p$ ($1-p$). It was already studied numerically
\cite{albano1,kolakowska2006} and with scaling approaches \cite{lam,rdcor,rdcoralbano}.
For small $p$, there is an enhancement of characteristic times of the correlated
kinetics ($p=1$) by a factor $p^{-1}$ and of the outer surface roughness
by a factor $p^{-1/2}$;
for a recent discussion of this topic, see Ref. \protect\cite{anomalouscompetitive}.
These features extend to other ballistic-like models with crossovers to UD
\cite{bbdflavio} and are related to the lateral aggregation.
In case of surface relaxation after aggregation,
the exponents in those relations are larger, corresponding to longer crossovers
for small $p$ \cite{rdcor,rdcoralbano,juvenil}.

A renewed interest in these competitive models was recently observed, with a focus on the
limitations of scaling relations or with an emphasis on the properties of porous media.
Refs. \protect\cite{kolakowskacomment,albanoreply,kolakowska2015} discussed the deviations
from the dominant scaling of surface roughness at low $p$, which is essential for
a quantitative characterization of surface properties in a variety of growth conditions.
The effect of relaxation after collision of incident and aggregated particles was
considered in Ref. \protect\cite{juvenil}, also with a focus on surface properties.
Refs. \protect\cite{banerjee,mal} considered the effect of a stickness
parameter on the aggregation of the incident particles, which may attach to neighboring
particles located below the outer surface. Simulations 
in $1+1$ dimensions produced deposits with porosity ranging from very small values to
approximately $70\%$ and suggested non-KPZ behavior in one of the models
\cite{banerjee}.

The first aim of this paper is to study surface and bulk properties of the model
proposed in Ref. \protect\cite{banerjee} combining a systematic analysis of simulation data
and a scaling approach for small values of
the stickness parameter. From the extrapolation of saturation roughness and
relaxation times, we show that the model has KPZ exponents in $1+1$ dimensions.
Comparison of roughness distributions confirms KPZ scaling in $2+1$ dimensions,
thus ruling out the proposal of non-KPZ exponents. 
In the limit of small stickness parameter $a$, the scaling approach
shows that the crossover time and the roughness scale as $a^{-2/3}$ and $a^{-1/3}$,
respectively, for all substrate dimensions.
These results show a shortened crossover when compared to all previously studied models with
an UD component \cite{rdcor,rdcoralbano,juvenil}, which is a consequence of
subsurface aggregation.
The same approach predicts porosity and pore height scaling as $a^{1/3}$ and
$a^{-1/3}$, respectively. These predictions will be confirmed numerically in $1+1$ and $2+1$
dimensions. The approach can be extended to the model introduced
in Ref. \protect\cite{mal}, with the same crossover exponents due to the similar
subsurface aggregation conditions.

The rest of this work is organized as follows. In Sec. \ref{model}, we present the
sticky particle deposition model. In Sec. \ref{roughness}, we analyze the surface roughness
scaling of simulated deposits in $1+1$ dimensions. In Sec. \ref{scaling}, we present a scaling
approach that relates surface properties to the stickness parameter, and confirm those
predictions with numerical simulations. In Sec. \ref{porosity},
we extend the scaling approach to relate the porosity and the average pore height with
the stickness parameter, again with support from numerical simulations.
In Sec. \ref{SPD3d}, we show that the main results extend to $2+1$ dimensions.
In Sec. \ref{conclusion}, we summarize our results and conclusions.

\section{Sticky particle deposition model}
\label{model}

In all models discussed in this paper, square particles of size $c$ are sequentially
released on randomly chosen columns of a one-dimensional discretized substrate of
lateral size $Lc$ and move in a direction perpendicular to the substrate.
Here, $L$ is the number of columns.
The time interval for deposition of one layer of atoms ($L$ atoms) is $\tau$.
Thus, at time $t$, the number of deposited layers is $t/\tau$.

The model proposed in Ref. \protect\cite{banerjee} is hereafter called sticky particle
deposition (SPD). In each site of the trajectory of the incident particle, it interacts with
particles in nearest neighbor (NN) sites at the same layer (same height above the deposit)
and particles in next neareast neighbor (NNN) sites at the layer immediately below it.
This interaction is represented by a probabilistic rule of aggregation at its current position.

The probability of aggregation to each neighbor is
\begin{equation}
p_{ag}=\frac{a}{{\left( r/c\right)}^n} ,
\label{pag}
\end{equation}
where $a$ is the stickness parameter, $r$ is the distance between the centers of the particles
($r/c=1$ for NN, $r/c=\sqrt{2}$ for NNN), and $n$ is an exponent related to the nature of the
interaction.
In Ref. \protect\cite{banerjee}, the cases $n=2$ and $n=6$ were respectively called
Coulomb-type and van der Walls-type interactions, with most results being presented for the former.
Here we will restrict the analysis to the case $n=2$, in which aggregation to NN and NNN have
probabilities $a$ and $a/2$, respectively.

\begin{figure}
\includegraphics[width=7cm]{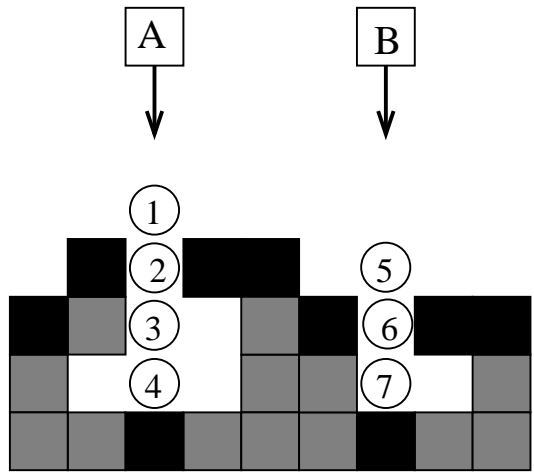}
\caption{Illustration of the rules of particle aggregation in the SPD model.
Deposited particles are gray and black squares, the latter indicating those particles in the outer surface.
Incident particles are indicated as squares marked A and B and circles numbered from 1 to 7 indicate
possible aggregation positions of those particles.
}
\label{modelSPD}
\end{figure}

Fig. \ref{modelSPD} helps to understand the aggregation rules of the SPD model and the
differences from other ballistic-like models. We first recall the rules
of BD and of the next nearest neighbor BD (BDNNN) model \cite{katzav2004,hivert,sigma}.

In BD, aggregation occurs at the first contact with a NN occupied site:
particle A at position 2, particle B at position 6 in Fig. \ref{modelSPD}.
In BDNNN, aggregation occurs at the first contact with a NNN occupied site:
position 1 for particle A and position 5 for particle B in Fig. \ref{modelSPD}.
In both cases, the incident particle interacts only with the
particles at the top of each column, which are highlighted in \ref{modelSPD}.
The set of top particles is called the outer surface of the deposit.

In the SPD model, particle A may aggregate at lattice sites marked with circles labeled 1 to 4,
and particle B may aggregate at lattice sites marked with circles labeled 5 to 7.

First, consider the trajectory of particle A.
In position 1, two aggregation trials are executed due to the interaction with two NNN
occupied sites; the probability of aggregation in each trial is $a/2$ .
If it does not aggregate there, it moves to the position labeled 2, in which three
aggregation trials are executed: two for interactions with the NN at the same height
(probability $a$ for each one) and one for interaction with the NNN in the layer below,
at the left (probability $a/2$).
If the particle does not aggregate at position 2, then it moves
to position 3 and may aggregate there with probability $a$ due to the interaction with the
NN at the left. If aggregation does not occur in position 3, the incident particle will
permanently aggregate at position 4, which is the top of the incidence column.
Relaxation to neighboring columns is not allowed.

Now we consider the trajectory of particle B.
In position 5, two aggregation trials are executed, each one with probability $a/2$
(due to interactions with two occupied NNN sites). If the particle does not aggregate there,
it moves to position 6, in which three aggregation trials are executed:
two for the interactions with the lateral NN (probability $a$ for
each trial) and one for the interaction with the NNN in the layer below (probability $a/2$).
If no aggregation trial
is successful at position 6, the particle moves to position 7 and aggregates there.

In contrast to other ballistic-like models (e. g. BD and BDNNN), the SPD model allows
subsurface aggregation. In Fig. \ref{modelSPD}, position 3 is an example of subsurface
position: it is not allowed in BD, nor in BDNNN, nor in any model of solid-on-solid
deposition (which prescribe aggregation at the top of each column).

In all cases, note that the interaction of an incident particle with an aggregated one is
possible in two steps: the first one when they are NNN (larger distance), the second one when
they are NN (smaller distance).
It represents two possibilities of aggregation in the ingoing part of the trajectory of
the incident particle.
If the aggregation trials are not accepted, then the incident particle moves to a lower
position. In this situation, this particle is in an outgoing trajectory respectively to those
aggregated particles.
For this reason, no aggregation trial is executed with a NNN aggregated particle
in the layer above the current position of the incident particle.
For instance, when particle A is at position 3 (third layer of the deposit), we do not
execute aggregation trials with the black NNN sites at the fourth layer in Fig. \ref{modelSPD}.

The SPD model resembles the model introduced in Ref. \protect\cite{gupta} and
the slippery BD model (SBD) proposed in Ref. \protect\cite{robledo}, both
studied in three-dimensional deposits (the latter with line seeds perpendicular
to a flat inactive surface).
Most of our simulation work is in $1+1$ dimensions, similarly to Ref. 
\protect\cite{banerjee}, but in Sec. \ref{SPD3d} we show that the main results
are also valid in three-dimensional samples.

For simplicity, in the following sections we consider unit values of the lattice constant
and of the time of deposition of a layer: $c=1$, $\tau =1$.

\section{Kinetic roughening}
\label{roughness}

The outer surface roughness is defined as
\begin{equation}
W(L,t)\equiv {\left[ { \left<  {\left( h - \overline{h}\right) }^2  \right> }
\right] }^{1/2} ,
\label{defw}
\end{equation}
where $h$ is the height of the top particle of each column,
the overbars indicate spatial averages, and the angular brackets indicate
configurational averages.
In systems with normal (in opposition to anomalous) scaling, the roughness follows
Family-Vicsek (FV) scaling \cite{fvpaper} as
\begin{equation}
W(L,t) \approx L^{\alpha} f\left( \frac{t}{t_\times}\right) ,
\label{fv}
\end{equation}
where $\alpha$ is the roughness exponent, $t_\times$ is a relaxation time,
and $f$ is a scaling function.
In long times ($t\gg t_\times$), $f\to const$, so that $W$ saturates as
\begin{equation}
W_{sat}\approx AL^{\alpha} ,
\label{wsat}
\end{equation}
where $A$ is a model-dependent constant. The saturation time
$t_\times$ scales as 
\begin{equation}
t_\times \approx BL^{z} , 
\label{scalingttimes}
\end{equation}
where $z$ is the dynamic exponent and $B$ is another model-dependent constant.
The roughness for $t\ll t_\times$ scales as
\begin{equation}
W \approx Ct^{\beta} ,
\label{defbeta}
\end{equation}
with $\beta = \alpha/z$ and another model-dependent constant $C$.

\begin{figure}
\includegraphics[width=7cm]{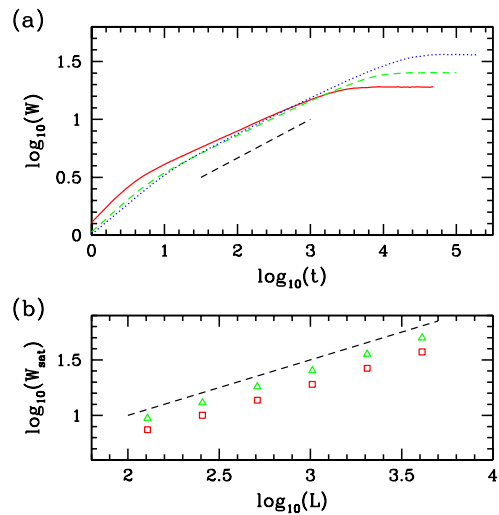}
\caption{(Color online) (a) Time evolution of the surface roughness in the SPD model
with $a=0.5$ (red solid curve), $a=0.1$ (green dashed curve),  and
$0.025$ (blue dotted curve). The dashed line has slope $1/3$ of KPZ scaling.
(b) Saturation roughness as a function of the lattice size for $a=0.5$ (red squares) and
$0.1$ (green triangles). The dashed line has slope $1/2$ of KPZ scaling.
}
\label{figroughness}
\end{figure}

Fig. \ref{figroughness}a shows the surface roughness evolution of the SPD model
for three values of $a$ in $L=1024$.

For short times, there is a crossover from an initial regime of rapid roughness increase to
a second regime in which it increases slower. For small $a$, the first regime is
mainly of UD and the slope of the $\log{W}\times \log{t}$ plot is near $1/2$.
For $a>0.1$, lateral aggregation is frequent, thus the roughness at short times is
larger than that of UD (e. g.  $a=0.5$ in Fig. \ref{figroughness}a).
It is difficult to find a pure UD regime in this case and to
estimate the crossover time with accuracy.

After this transient, the growth regime begins, with apparent power law scaling of $W$
[Eq. (\ref{defbeta})].
It is difficult to distinguish the different curves for small $a$ in Fig. \ref{figroughness}a;
this will be explained by the scaling approach of Sec. \ref{scaling}.
The slope of those curves are near $1/3$, suggesting KPZ scaling.

At long times, there is an increase in the saturation roughness as $a$ decreases.

Fig. \ref{figroughness}b shows the saturation roughness as a function of lattice size $L$
for two values of $a$. They seem to be consistent with the KPZ exponent $\alpha = 0.5$.
However, linear fits of those plots give slopes slightly smaller than $0.5$,
similarly to what was found in Ref. \protect\cite{banerjee}. For this reason, a systematic
extrapolation of those results is necessary to decide whether the roughness scaling
is KPZ or not.
We proceed by using the same methods of Refs. \protect\cite{balfdaar,tau,bbdflavio,sigma},
in which roughness scaling of various ballistic-like models was studied.

Effective roughness exponents are defined as
\begin{equation}
\alpha_L \equiv \frac{\ln \left[W_{sat}\left( L\right) /W_{sat}\left( L/2\right)\right]}{\ln{2}} .
\label{defalphaL}
\end{equation}
Assuming that the saturation roughness has scaling corrections as
$W_{sat}\sim L^{\alpha} (a_0+a_1 L^{-\Delta})$ \cite{balfdaar},
where $a_0$ and $a_1$ are constants, we expect
$\alpha_L \approx \alpha + a_2 L^{-\Delta}$, where $a_2$ is another constant.

\begin{figure}
\includegraphics[width=7cm]{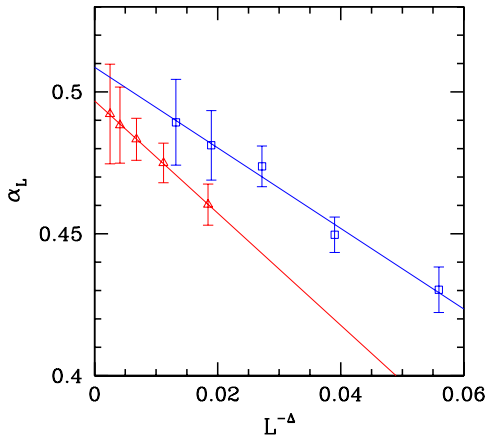}
\caption{(Color online) Effective roughness exponents as a function of $L^{-\Delta}$ for
$a=0.1$ (red triangles) with $\Delta=0.72$ and $a=0.5$ (blue squares) with $\Delta=0.52$.
}
\label{alpha}
\end{figure}

Fig. \ref{alpha} shows effective exponents as a function of $L^{-\Delta}$
for $a=0.5$ and $a=0.1$, respectively using $\Delta =0.52$ and
$\Delta=0.72$. These values of $\Delta$
provide the best linear fits of the $\alpha_L\times L^{-\Delta}$ data for
each stickness parameter.
The asymptotic ($L\to\infty$) estimates from those fits are $\alpha=0.496\pm 0.015$ and
$\alpha=0.51\pm 0.02$, respectively.

We estimate the dynamical exponent $z$ using the method proposed in Ref. \protect\cite{tau}.
For each lattice size $L$, a characteristic time $t_0$ is defined as
\begin{equation}
W\left( t_0\right) = k {W_{sat}} ,
\label{deft0}
\end{equation}
with $k<1$.
The FV relation (\ref{fv}) shows that $t_0$ is proportional to $t_\times$ for fixed $k$,
thus $t_0\sim L^z$. Effective dynamical exponents are defined as
\begin{equation}
z_L \equiv \frac{\ln \left[t_0\left( L\right) /t_0\left( L/2\right)\right]}{\ln{2}} .
\label{defzL}
\end{equation}

\begin{figure}
\includegraphics[width=7cm]{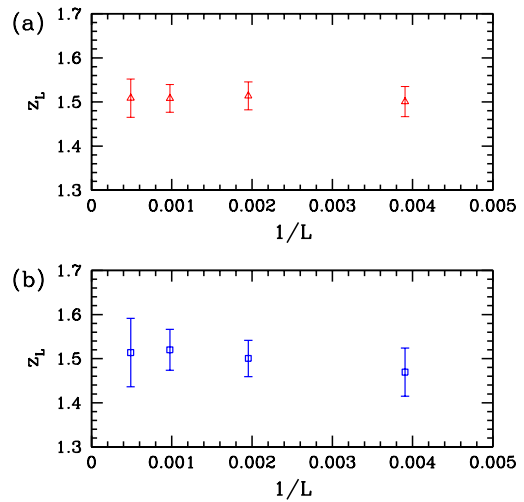}
\caption{(Color online) Effective dynamical exponents as a function of $1/L$ for
(a) $a=0.1$ and (b) $a=0.5$.}
\label{zL}
\end{figure}

Figs. \ref{zL}a and \ref{zL}b show $z_L$ for $a=0.5$ and $a=0.1$,
respectively, obtained with $k=0.8$. In both cases, the exponents oscillate near $z=1.5$,
suggesting that finite-size corrections are very small. 

The estimates of $\alpha$ and $z$ are in very good agreement with
KPZ exponents $\alpha=1/2$ and $z=3/2$, which is a strong numerical evidence that this model
is in the KPZ class in $1+1$ dimensions.

A universal scaling is expected in the SPD model because there is no change in its
symmetries as the stickness parameter changes. In other words, the corresponding hydrodynamic
growth equation may have coefficients dependent on the parameter $a$, but the leading
spatial derivatives will be the same \cite{barabasi,hagston}. 
Due to the lateral aggregation and consequent excess growth velocity,
KPZ scaling is expected for any $a>0$.

Previous works on ballistic-like models
\cite{balfdaar,sigma,farnudiPRE,farnudiJPCS,balrsos2013,sidiney2014} 
have already shown that systematic extrapolation of finite-size or finite-time data
are necessary to avoid crossover effects. 
As highlighted in Ref. \protect\cite{sidiney2014}, this is a consequence of
the large fluctuations in height increments, typical of those models.

Crossovers and finite-size corrections probably are the reasons for the deviations from
KPZ scaling observed in Ref. \protect\cite{banerjee}. This may also be inferred by comparison
with finite-size BD data from Ref. \protect\cite{balfdaar}.
Ref. \protect\cite{banerjee} suggests $\alpha\approx 0.42$ for $a=1$,
while Ref. \protect\cite{balfdaar} gives effective exponents $0.40 \leq \alpha_L\leq 0.45$
for BD in the same range of $L$.
Moreover, the growth exponents $\beta$ in lattice sizes from $L=256$ to $L=1024$
for $a=1$, shown in Ref. \protect\cite{banerjee},
are very near the corresponding estimates for BD in
Ref. \protect\cite{balfdaar} (considering minimum linear correlation coefficient $0.999$ in
the growth region).

\section{Scaling for small stickness parameter}
\label{scaling}

\subsection{Scaling approach}
\label{approach}

For small $a$, lateral aggregation is unprobable, thus most particles aggregate at the top of
the column of incidence. At short times, the roughness increases as \cite{barabasi}
\begin{equation}
W_{RD}\approx t^{1/2} .
\label{wrandom}
\end{equation}
After a crossover time $t_c$, KPZ scaling appears.
Our first step is to relate $t_c$ to $a$, for $a\ll 1$.

\begin{figure}
\includegraphics[width=2.5cm]{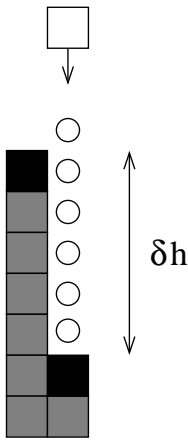}
\caption{Two neighboring columns after some time of random growth.
Deposited particles are gray and black squares, the latter indicating particles at
the outer surface. The incident particle is the empty square and circles
indicate possible aggregation positions of this particle. }
\label{columns}
\end{figure}

A typical configuration of two neighboring columns in UD is illustrated in Fig. \ref{columns}.
It has a height difference
\begin{equation}
\delta h\sim t^{1/2} 
\label{deltah}
\end{equation}
because their heights increase without correlations.
If $\delta h$ is large, then a new particle inciding at the right column in Fig. \ref{columns}
may aggregate at a number of positions of order $\delta h$, as indicated by the circles.
This means that the number of aggregation trials is of order $\delta h$ and
the aggregation probability for each trial is of order $a$.
Thus, the probability of no lateral aggregation after those trials
(i. e. aggregation at the top of the column)
is $P_{top}\sim {\left( 1-a\right)}^{\delta h}$.
The probability that some lateral aggregation occurs is, consequently,
\begin{equation}
P_{lat} = 1-P_{top} \sim 1-{\left( 1-a\right)}^{\delta h}\approx a\delta h .
\label{Plat}
\end{equation}
The latter approximation requires $P_{lat}\sim a\delta h \ll 1$, which will be confirmed below.

The average time for a lateral aggregation event at a given column is
$t_{lat}\sim 1/P_{lat}$.
Lateral aggregation immediately creates correlations between the heights of neighboring columns,
thus the crossover time is $t_c\approx t_{lat}$.
Eqs. (\ref{wrandom}) and (\ref{Plat}) at the crossover
($\delta h = {\delta h}_c$, $t=t_c$) give
\begin{equation}
{\delta h}_c \sim {t_c}^{1/2} \sim {\left(\frac{1}{P_{lat}}\right)}^{1/2}
\sim {\left( a{\delta h}_c\right)}^{-1/2} .
\label{deltah_tc}
\end{equation}
Thus, height fluctuations at the crossover scale as 
\begin{equation}
{\delta h}_c \sim a^{-1/3}
\label{deltahc}
\end{equation}
and the crossover time scales as
\begin{equation}
t_c \sim a^{-2/3} .
\label{tc}
\end{equation}
These results confirm that $P_{lat}\sim a\delta h \leq a{\delta h}_c\ll 1$,
as the approximation in Eq. (\ref{Plat}) requires.

The amplitudes of the saturation roughness [Eq. (\ref{fv})] and of the relaxation time
[Eq. (\ref{scalingttimes})] scale as ${\delta h}_c$ and $t_c$, respectively.
Following the exponent convention introduced by Horowitz and Albano \cite{albano1,albano2},
we have
\begin{equation}
A\sim a^{-\delta} \qquad , \qquad \delta=1/3 ,
\label{defdelta}
\end{equation}
and
\begin{equation}
B\sim a^{-y} \qquad , \qquad y=2/3 .
\label{defy}
\end{equation}
These results are valid in any spatial dimension because UD
properties are not dimension-dependent.

The scaling exponents in Eqs. (\ref{defdelta}) and (\ref{defy}) differ from those obtained
in other competitive models with ballistic-like aggregation with probability $p$ and UD 
with probability $1-p$; in those systems, $\delta =1/2$ and $y=1$
\cite{albano1,rdcor,bbdflavio}. In solid-on-solid models with crossovers from UD to
correlated growth, the exponents are also different: $\delta=1$ and $y=2$.
The shorter crossover of the SPD model is due to subsurface aggregation, which
provides a large number of opportunities (of order ${\delta h}_c$)
for lateral aggregation of the incident particle (Fig. \ref{columns}).
On the other hand, the relation $\delta =y/2$ obtained in other competitive models is also
obeyed here because it is solely related to UD scaling (see e. g. the discussion in
Ref. \protect\cite{anomalouscompetitive}).

\subsection{Numerical test}

We performed simulations of the SPD in $L=1024$ and small values of $a$,
from $0.1$ to $0.0125$, until the steady states (roughness saturation).
The saturation roughness $W_{sat}$ and the characteristic times $t_0$ 
were calculated following the same lines of Sec. \ref{roughness}.

\begin{figure}
\includegraphics[width=7cm]{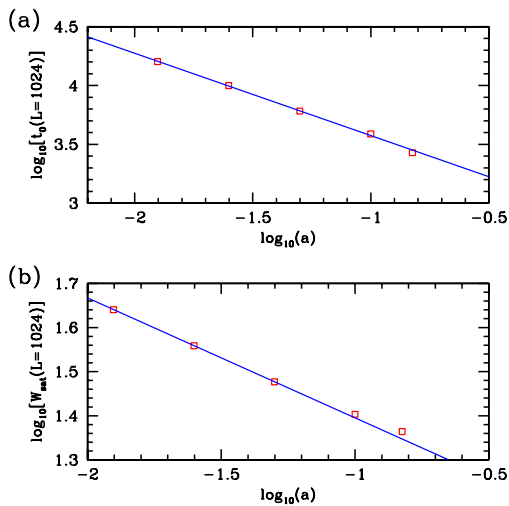}
\caption{(Color online) (a) Characteristic time $t_0$ and (b) saturation roughness $W_{sat}$
in size $L=1024$ as a function of the stickness parameter. Solid lines are fits of the data
for $a\leq 0.05$.}
\label{wt}
\end{figure}

Figs. \ref{wt}a and \ref{wt}b show $t_0$ and $W_{sat}$, respectively, as a function of
the stickness parameter $a$.
Since they were measured for constant $L$, they are expected to scale as the amplitudes
$B$ [Eq. (\ref{scalingttimes})] and $A$ [Eq. (\ref{wsat})], respectively.
Fits of the data for $a\leq 0.1$ give exponents $y\approx 0.70$ [Eq. (\ref{defy})
and Fig. \ref{wt}a] and $\delta\approx 0.27$ [Eq. (\ref{defdelta}) and Fig. \ref{wt}b].

The estimate of $y$ is in good agreement with the theoretical prediction of Eq. (\ref{defy}).
However, the estimate of $\delta$ shows a discrepancy of $\approx 20\%$
from the theoretical prediction of Eq. (\ref{defdelta}).
Note that the fits in Figs. \ref{wt}a and \ref{wt}b considered $0.05\leq a\geq 0.0125$,
which are not very small values of $a$, thus deviations are expected, particularly
in the smaller exponent ($\delta$).
Unfortunately, it is very difficult to obtain accurate estimates for smaller values of $a$
because relaxation times become very large and roughness fluctuations also increase.
Using smaller system sizes is also inappropriate because it enhances crossover effects.

Ref. \protect\cite{banerjee} estimated the crossover times for $L=512$ and obtained
$t_c\sim a^{-0.4\pm 0.04}$, which is significanly different from the theoretical prediction
in Eq. (\ref{tc}). However, measuring reliable crossover times is a difficult task, as 
explained in Sec. \ref{roughness}.
On the other hand, the same work shows that the saturation time for $L=1024$ scales as
$a^{-0.7\pm 0.03}$, which is in good agreement with our estimate.

The scaling of the amplitude $C$ in Eq. (\ref{defbeta}) can be predicted along the
same lines of Refs. \protect\cite{albano1,rdcor,rdcoralbano} for other competitive models:
\begin{equation}
C\sim a^{\gamma} \qquad ,\qquad \gamma =\delta -y\beta .
\label{defgamma}
\end{equation}
For the SPD model in $1+1$ dimensions, we obtain $\gamma = 1/9$.
This very small exponent gives a very slow variation of $C$ with the stickness parameter.
It explains the small distance between the curves for different values of $a$ in
Fig. \ref{figroughness}a.

\subsection{Scaling in a related model}

In Ref. \protect\cite{mal}, a model similar to the SPD was introduced. The particles
incide vertically and, at each site of its trajectory with a NN occupied site,
it may aggregate with probability $p$. Otherwise, the particle moves down one site.
If no lateral aggregation occurs, the particle aggregates at the top of the column
of incidence.

In Fig. \ref{modelSPD}, particle A may aggregate to positions labeled 2, 3, and 4.
In positions 2 and 3, aggregation trials have probability $p$.
If the particle does not aggregate at one of those points, 
it moves to position 4 and aggregates there.
Particle B may aggregate at position 6 with
probability $p$, otherwise it moves to position 7 and aggregates there.

For small $p$, most lateral aggregation trials are rejected, thus UD dominates.
Large local height fluctuations appear, similarly to Fig. \ref{columns}.
The increase of the local height difference $\delta h$ and the probability of
lateral aggregation $P_{lat}$ are given by Eqs. (\ref{deltah}) and (\ref{Plat}),
with $a$ replaced by $p$. Thus, the same reasoning of Sec. \ref{approach} leads to the
same scaling relations of the SPD model with $a$ replaced by $p$.

In the notation of Ref. \protect\cite{mal}, exponents $\alpha '=1/3$ and
$z'=2/3$ are predicted by our scaling approach. The numerical estimates of
that work, $\alpha '\approx 0.25$ and $z'\approx 0.77$, differ from those
predictions, probably because they were obtained by data collapse methods
that do not account for scaling corrections.

\section{Porosity and pore height scaling}
\label{porosity}

\begin{figure}
\includegraphics[width=7cm]{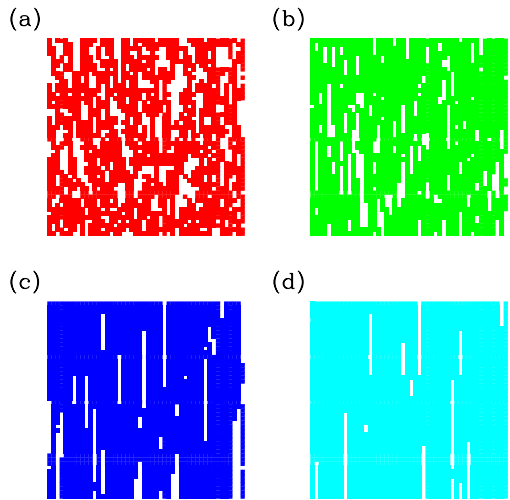}
\caption{(Color online) Regions of size $48\times 48$ (in lattice units) of samples grown
with stickness parameters (a) $0.1$, (b) $0.01$, (c) $0.001$, and (d) $0.0001$.}
\label{samplesSPD}
\end{figure}

For $a=1$, the samples have large porosity $\Phi\approx 70\%$.
When $a$ decreases, $\Phi$ decreases because UD creates
no holes. Fig. \ref{samplesSPD} shows regions of some samples obtained
with small stickness parameters. The porosity decrease is accompanied by the
formation of longer pores extended in the vertical direction. This is a consequence
of the increase of the height fluctuation ${\delta h}_c$ before a lateral aggregation event
[Eq. (\ref{deltahc})].

The number of deposited layers necessary to attain a steady state value of $\Phi$ is
relatively small, typically of the order of $t_c$ in Eq. (\ref{tc}).
This is expected because pores are narrow, even in pure BD, thus porosity depends
only on short wavelength height fluctuations, which saturate at short times
(in the absence of scaling anomaly \cite{ramasco}).

Our scaling approach can be used to predict the dependence of the porosity $\Phi$
and the average pore height on the parameter $a$, as follows.

During the time interval $t_c$ between two lateral aggregation events, the number of
particles deposited at the top of a given column is approximately $t_c$
(note that we are still using unit lattice constant and unit deposition time of a layer,
$c=1$ and $\tau =1$).
The size of a long pore produced by the lateral aggregation is ${\delta h}_c$
[Eq. (\ref{deltahc})]. Consequently, for small $a$, the porosity (pore volume divided
by total volume) is expected to scale as
\begin{equation}
\Phi \sim \frac{{\delta h}_c}{t_c+{\delta h}_c} \sim a^{1/3} .
\label{Pscaling}
\end{equation}
This is valid in the limit of very small $a$, in which ${\delta h}_c \ll t_c$.

The small exponent in Eq. (\ref{Pscaling}) explains why a large decrease of $a$ leads
only to mild reduction of porosity.
This is remarkably illustrated in Fig. \ref{samplesSPD}, in which $a$ varies three orders
of magnitude, while the porosity decreases from $\Phi\approx 0.39$ to $0.044$, i. e.
changes by a factor smaller than $10$.

The porosity scaling in the SPD also differs from other competitive models involving
ballistic-type aggregation. Examples are the bidisperse ballistic deposition \cite{bbdflavio}
and the BD-UD competitive model, in which $\Phi\sim p^{1/2}$ ($p$ is
the probability of the ballistic-like component).

We simulated the SPD in size $L=1024$ for small values of $a$ in order to measure
the porosity between times $t_I=5000$ and $t_F=10000$. In all cases,
$t_I$ is much larger than the crossover time and
$t_F$ is much smaller than the relaxation time $t_\times$.

\begin{figure}
\includegraphics[width=7cm]{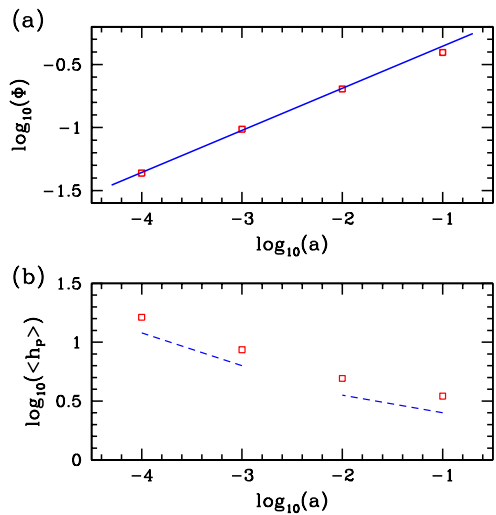}
\caption{(Color online) (a) Porosity as a function of the stickness parameter. The solid line
is a linear fit of the data for $a\leq 0.01$. (b) Average pore height as a function of the
stickness parameter. Dashed lines have slopes $-0.15$ (right) and $-0.28$ (left).}
\label{Phi}
\end{figure}

Fig. \ref{Phi}a shows the porosity as a function of the stickness parameter.
The linear fit for ${10}^{-4}\leq a\leq {10}^{-2}$ gives $\Phi\sim a^{-0.33}$, in excellent
agreement with Eq. (\ref{Pscaling}). Although these values of $a$ are very small, 
the corresponding values of $a^{1/3}$ and of $\Phi$ are not very small.
Thus, scaling corrections are particularly weak in this case.

For very small $a$, the pores are long and isolated, as illustrated in Fig. \ref{samplesSPD}d.
The average pore height is expected to scale as Eq. (\ref{deltahc}), because a pore is
formed only when a lateral aggregation event occurs.
However, for $\Phi\sim 0.1$ or larger, many pores occupy two
or more neighboring columns. This can be observed in
Figs. \ref{samplesSPD}a, \ref{samplesSPD}b, and \ref{samplesSPD}c.

Here we define pore height as the vertical distance between the 
aggregation position and the top of the incidence column in any lateral aggregation event.
Its average value, $\langle h_P\rangle$, is taken over all lateral aggregation events
between $t_I$ and $t_F$ in ${10}^3$ different samples. 
For small $\Phi$, pores are isolated, thus $\langle h_P\rangle$ is a reliable
approximation of the average pore height, and is expected to scale as Eq. (\ref{deltahc}).
For $\Phi$ not too small, some pores occupy two or more neighboring columns,
and all these columns contribute to $\langle h_P\rangle$ (each one had a lateral
aggregation event).

Fig. \ref{Phi}b shows $\langle h_P\rangle$ as a function of $a$.
The slope of that log-log plot evolves from $-0.15$ for ${10}^{-1}\leq a\leq {10}^{-2}$
to $-0.28$ for ${10}^{-3}\leq a\leq {10}^{-4}$.
The latter is $16\%$ smaller than the theoretically predicted value $-1/3$
[Eq. (\ref{deltahc})], which
indicates the presence of large scaling corrections.

Ref. \protect\cite{banerjee} measured the porosity of samples with $0.1\leq a\leq 1$,
with results in qualitative agreement with ours. However, the low porosity scaling
was not addressed there.

Ref. \protect\cite{mal} suggests that the porosity scales with $p$ (equivalent to $a$)
and with the lattice size $L$. The latter is expected only as vanishing corrections,
since porosity does not depend on long wavelength fluctuations.
This explains the small (effective) exponents $a$ and
$c$ obtained in that work. On the other hand, Ref. \protect\cite{mal} estimates
the long-time scaling on $p$ with exponent $b=0.22$, which is to be compared with the
theoretical prediction $1/3$. The discrepancy is probably related to
the use of data collapse methods.

\section{SPD model in $2+1$ dimensions}
\label{SPD3d}

The aim of this section is to show that the main features of the SPD model in $1+1$ dimensions
can be extended to $2+1$ dimensions, namely the KPZ roughening of the outer surface and the
porosity scaling derived by the superuniversal approach of Sec. \ref{scaling}.

The aggregation rules of the SPD model have to be extended in this case.
First, NN interactions are considered in two substrate directions, with a total of four
NN in the same height.
Secondly, NNN interactions appear with aggregated particles in the same height
(four neighbors) and with particles at the level immediately below (four neighbors).

Roughness scaling of ballistic-like models usually
show large corrections \cite{balfdaar,sigma,bbdflavio}. An alternative
to search for the universality class of a given model is the comparison of scaled
roughness distributions of relatively small systems because the finite-size corrections
of those quantities are much smaller \cite{sigma,bbdflavio}.

We simulated the SPD model with $a=0.1$ in substrates of lateral size $L=256$ up to
the steady state (roughness saturation). In this regime, the square roughness
$w_2\equiv \overline{h^2} - {\overline{h}}^2$ of several configurations is measured.
$P\left( w_2\right)$ is the probability density of the square roughness of a given
configuration to lie in the range $\left[ w_2, w_2+dw_2\right]$.
This quantity is expected to scale as
\begin{equation}
P\left( w_2\right) = {1\over \sigma}
\Psi\left( {{w_2-\left< w_2\right>}\over\sigma}\right) ,
\label{scalingdist}
\end{equation}
where $\sigma \equiv \sqrt{ \left< {w_2}^2 \right> - {\left< w_2\right>}^2 }$ is
the rms fluctuation of $w_2$ and $\Psi$ is a universal function \cite{foltin,racz,antal}.

\begin{figure}
\includegraphics[width=7cm]{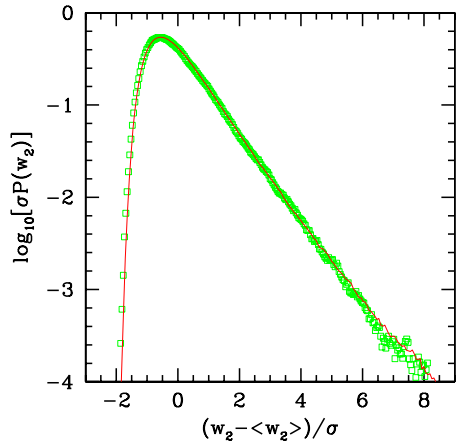}
\caption{(Color online) Scaled roughness distributions in the steady states of the SPD
model with $a=0.1$ (squares) and of the RSOS model (solid curve) in $2+1$ dimensions,
with $L=256$.
}
\label{dist}
\end{figure}

Fig. \ref{dist} shows the scaled roughness distribution of the SPD model and
the distribution of the restricted solid-on-solid (RSOS) model \cite{kk} in substrate
size $L=256$.
The latter is a well known representative of the KPZ class and its roughness distributions
have negligible finite-size effects \cite{distrib}. The excellent collapse of the
curves in Fig. \ref{dist} is striking evidence that the SPD model also belongs to the
KPZ class in $2+1$ dimensions.

We also simulated the SPD model in size $L=1024$ for small values of $a$ and measured
the porosity between times $t_I=1000$ and $t_F=2000$. We observe that the porosity is
larger than in the $1+1$-dimensional samples for the same value of $a$. For instance,
for $a=0.1$, the porosity exceeds $50\%$.
This is a consequence of the larger number of interactions of the incident particle
with NN and NNN in $2+1$ dimensions, which facilitates lateral aggregation.

\begin{figure}
\includegraphics[width=7cm]{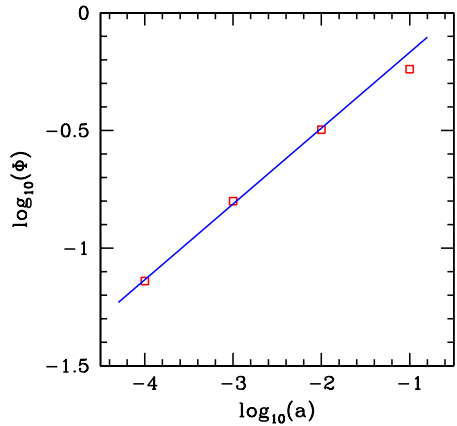}
\caption{(Color online) Porosity as a function of the stickness parameter of the SPD model
in $2+1$ dimensions. The solid line is a linear fit of the data for $a\leq 0.01$.
}
\label{Phi3d}
\end{figure}

Fig. \ref{Phi3d} shows the porosity as a function of $a$, for low values of that parameter.
The linear fit for ${10}^{-4}\leq a\leq {10}^{-2}$ gives $\Phi\sim a^{-0.32}$, which is also
in excellent agreement with Eq. (\ref{Pscaling}). This supports the extension of the
scaling approach of Sec. \ref{scaling} to $2+1$ dimensions.

An important consequence of this scaling approach is to facilitate the design of
samples with the desired values of porosity and elongate pores.
However, one has to take care with the fluctuations in the value of $\Phi$ in
the first layers of the deposit, typically produced at $t\lesssim t_c$.

\section{Conclusion}
\label{conclusion}

We studied surface and bulk properties of porous deposits produced by a
model proposed in Ref. \protect\cite{banerjee}, in substrates with one and two
dimensions. The model shows a crossover from uncorrelated to correlated growth
for small values of the stickness parameter $a$.

In $1+1$ dimensions, a systematic analysis of simulation data for saturation 
roughness and relaxation times shows that the model belongs to the KPZ class.
Finite-size corrections explain the previous claim of deviations from KPZ
scaling. In $2+1$ dimensions, KPZ roughening is confirmed by comparison of
roughness distributions.

A scaling approach for small values of $a$ is proposed to
relate the crossover time and the local height fluctuations with that parameter,
respectively giving exponents $-2/3$ and $-1/3$. These results are consequence
of the UD properties, thus they do not depend on the spatial dimension.
Numerical results confirm these predictions. The crossover exponents are smaller than
those of other competitive models that consider aggregation only at the outer
surface \cite{rdcor,rdcoralbano,juvenil}.

The same approach predicts the porosity scaling as $a^{1/3}$, which is
in good agreement with simulation results in $1+1$ and $2+1$ dimensions.
This result is important for using the model to produce porous samples
representative of real materials. This may also help to model
samples with desired porosity and pore height, particularly
for the possibility of controlling the scaling properties by changing
the kinetics of subsurface aggregation.

\acknowledgements

This work was partially supported by CNPq and FAPERJ (Brazilian agencies).


\end{document}